\documentclass[11pt]{article}
\input epsf
\usepackage{psfrag}
\usepackage{graphicx}

\catcode`\@=11
\def\@citex[#1]#2{%
\if@filesw \immediate \write \@auxout {\string \citation {#2}}\fi
\@tempcntb\m@ne \let\@h@ld\relax \def\@citea{}%
\@cite{%
  \@for \@citeb:=#2\do {%
    \@ifundefined {b@\@citeb}%
      {\@h@ld\@citea\@tempcntb\m@ne{\bf ?}%
      \@warning {Citation `\@citeb ' on page \thepage \space undefined}}%
      {\@tempcnta\@tempcntb \advance\@tempcnta\@ne%
      \@tempcntb\number\csname b@\@citeb \endcsname \relax%
      \ifnum\@tempcnta=\@tempcntb 
        \ifx\@h@ld\relax%
          \edef \@h@ld{\@citea\csname b@\@citeb\endcsname}%
        \else%
          \edef\@h@ld{\ifmmode{-}\else--\fi\csname b@\@citeb\endcsname}%
        \fi%
      \else
        \@h@ld\@citea\csname b@\@citeb \endcsname%
        \let\@h@ld\relax%
      \fi}%
    \def\@citea{,\penalty\@highpenalty\,}%
  }\@h@ld
}{#1}}

\def\@citeb#1#2{{[#1]\if@tempswa , #2\fi}}
\def\@citeu#1#2{{$^{#1}$\if@tempswa , #2\fi }}
\def\@citep#1#2{{#1\if@tempswa , #2\fi}}

\def\bcites{         
        \catcode`\@=11
        \let\@cite=\@citeb
        \catcode`\@=12
}

\def\upcites{         
        \catcode`\@=11
        \let\@cite=\@citeu
        \catcode`\@=12
}

\def\plaincites{      
        \catcode`\@=11
        \let\@cite=\@citep
        \catcode`\@=12
}

\newcount\hour
\newcount\minute
\newtoks\amorpm
\hour=\time\divide\hour by 60
\minute=\time{\multiply\hour by 60 \global\advance\minute by-\hour}
\edef\standardtime{{\ifnum\hour<12 \global\amorpm={am}%
        \else\global\amorpm={pm}\advance\hour by-12 \fi
        \ifnum\hour=0 \hour=12 \fi
        \number\hour:\ifnum\minute<10 0\fi\number\minute\the\amorpm}}
\edef\militarytime{\number\hour:\ifnum\minute<10 0\fi\number\minute}

\def\draftlabel#1{{\@bsphack\if@filesw {\let\thepage\relax
   \xdef\@gtempa{\write\@auxout{\string
      \newlabel{#1}{{\@currentlabel}{\thepage}}}}}\@gtempa
   \if@nobreak \ifvmode\nobreak\fi\fi\fi\@esphack}
        \gdef\@eqnlabel{#1}}
\def\@eqnlabel{}
\def\@vacuum{}
\def\marginnote#1{}
\def\draftmarginnote#1{\marginpar{\raggedright\scriptsize\tt#1}}
\def\draft{
        \pagestyle{plain}
        \overfullrule=2pt
        \oddsidemargin -.5truein
        \def\@oddhead{\sl \phantom{\today\quad\militarytime} \hfil
        \smash{\Large\sl DRAFT} \hfil \today\quad\militarytime}
        \let\@evenhead\@oddhead
        \let\label=\draftlabel
        \let\marginnote=\draftmarginnote
        \def\ps@empty{\let\@mkboth\@gobbletwo
        \def\@oddfoot{\hfil \smash{\Large\sl DRAFT} \hfil}
        \let\@evenfoot\@oddhead}
        \def\@eqnnum{(\theequation)\rlap{\kern\marginparsep\tt\@eqnlabel}%
        \global\let\@eqnlabel\@vacuum}  }

\def\blackfonts{
        \font\blackboard=msbm10 scaled\magstep1
        \font\blackboards=msbm8
        \font\blackboardss=msbm6
}

\def\prep{         
        \catcode`\@=11
        \input art10.sty
        \catcode`\@=12
        
        \let\small\null
        \def\blackfonts{
                \font\blackboard=msbm10
                \font\blackboards=msbm7
                \font\blackboardss=msbm5
        }
        \let\sl\it
        \twocolumn
        \sloppy
        \voffset=-2.54truecm
        \hoffset=-2.54truecm
        \flushbottom
        \parindent 1em
        \leftmargini 2em
        \leftmarginv .5em
        \leftmarginvi .5em
        \marginparwidth 48pt
        \marginparsep 10pt
        \setlength{\columnsep}{2truecm}
        \setlength{\textwidth}{25.4truecm}
        \setlength{\textheight}{17truecm}
        \baselineskip=16pt
        \oddsidemargin .18truein
        \evensidemargin .17truein
}

\def\eqalign#1{\null\,\vcenter{\openup\jot\m@th
  \ialign{\strut\hfil$\displaystyle{##}$&$\displaystyle{{}##}$\hfil
      \crcr#1\crcr}}\,}
\def\eqalignno#1{\displ@y \tabskip\centering
  \halign to\displaywidth{\hfil$\@lign\displaystyle{##}$\tabskip\z@skip
    &$\@lign\displaystyle{{}##}$\hfil\tabskip\centering
    &\llap{$\@lign##$}\tabskip\z@skip\crcr
    #1\crcr}}

\def\section{\@startsection {section}{1}{\z@}{3.ex plus 1ex minus
 .2ex}{2.ex plus .2ex}{\large\bf}}
\def\subsection{\@startsection{subsection}{2}{\z@}{2.75ex plus 1ex minus
 .2ex}{1.5ex plus .2ex}{\bf}}        

\def\appendix{{\newpage\section*{Appendix}}\let\appendix\section%
        {\setcounter{section}{0}
        \gdef\thesection{\Alph{section}}}\section}

\def\abstract{\if@twocolumn
\section*{Abstract}
\else 
\begin{center}
{\bf Abstract\vspace{-.5em}\vspace{0pt}}
\end{center}
\quotation
\fi}

\catcode`\@=12

\def\d{\partial}

\def\sqr#1#2{{\vcenter{\vbox{\hrule height.#2pt\hbox{\vrule width.#2pt 
height#1pt \kern#1pt \vrule width.#2pt}\hrule height.#2pt}}}}
 
\def\fii{\varphi}

\def\=d{\,{\buildrel\rm def\over =}\,}

\def\i3p{\p32\int d^3p}

\def\As{A\hbox to 1pt{\hss /}}
\def\np4{\int d^4p_1\cdots d^4p_{n-1}\, }

\def\nx4{\int d^4x_1\ldots d^4x_n\, }

\def\kon#1#2{\vbox{\halign{##&&##\cr
\lower4pt\hbox{$\scriptscriptstyle\vert$}\hrulefill &
\hrulefill\lower4pt\hbox{$\scriptscriptstyle\vert$}\cr $#1$&
$#2$\cr}}}

\def\konv#1#2#3{\hbox{\vrule height12pt depth-1pt}
\vbox{\hrule height12pt width#1cm depth-11.6pt}
\hbox{\vrule height6.5pt depth-0.5pt}
\vbox{\hrule height11pt width#2cm depth-10.6pt\kern5pt
      \hrule height6.5pt width#2cm depth-6.1pt}
\hbox{\vrule height12pt depth-1pt}
\vbox{\hrule height6.5pt width#3cm depth-6.1pt}
\hbox{\vrule height6.5pt depth-0.5pt}}
\def\konu#1#2#3{\hbox{\vrule height12pt depth-1pt}
\vbox{\hrule height1pt width#1cm depth-0.6pt}
\hbox{\vrule height12pt depth-6.5pt}
\vbox{\hrule height6pt width#2cm depth-5.6pt\kern5pt
      \hrule height1pt width#2cm depth-0.6pt}
\hbox{\vrule height12pt depth-6.5pt}
\vbox{\hrule height1pt width#3cm depth-0.6pt}
\hbox{\vrule height12pt depth-1pt}}

\def\konw#1#2#3{\hbox{\vrule height12pt depth-1pt}
\vbox{\hrule height12pt width#1cm depth-11.6pt}
\hbox{\vrule height6.5pt depth-0.5pt}
\vbox{\hrule height12pt width#2cm depth-11.6pt \kern5pt
      \hrule height6.5pt width#2cm depth-6.1pt}
\hbox{\vrule height6.5pt depth-0.5pt}
\vbox{\hrule height12pt width#3cm depth-11.6pt}
\hbox{\vrule height12pt depth-1pt}}

\def\i{{\rm int}}

\def\r{{\rm ret}}
\def\a{{\rm av}}

\def\m3{{\mu_1\mu_2\mu_3}}

\def\p{{(+)}}

\def\be{\begin{equation}}       \def\eq{\begin{equation}}
\def\ee{\end{equation}}         \def\eqe{\end{equation}}

\def\bea{\begin{eqnarray}}      \def\eqa{\begin{eqnarray}}
\def\ena{\end{eqnarray}}        \def\eea{\end{eqnarray}}
                                \def\eqae{\end{eqnarray}}

\def\ba{\begin{array}}
\def\ea{\end{array}}
\def\unit{1 \hskip-.3em \raise2pt\hbox{$ \scriptstyle |$ } }

\def\a{\alpha}
\def\b{\beta}
 
\def\d{\delta}
\def\g{\gamma}
   
\def\i{\iota}
\def\j{\psi}
\def\l{\lambda}
\def\m{\mu}
\def\n{\nu}
  
\def\p{\pi}                
\def\r{\rho}                                     
\def\s{\sigma}                                   
\def\t{\tau}

\def\G{\Gamma}

\def\car{{\cal R}}

\def\bop#1{\setbox0=\hbox{$#1M$}\mkern1.5mu
        \vbox{\hrule height0pt depth.04\ht0
        \hbox{\vrule width.04\ht0 height.9\ht0 \kern.9\ht0
        \vrule width.04\ht0}\hrule height.04\ht0}\mkern1.5mu}
\def\pa{\partial}                              

\def\>{\rangle} 

\def\<{\langle} 
\def\Dsl{D \hskip-.6em \raise1pt\hbox{$ / $ } }

\def\sl#1{\rlap{\hbox{$\mskip 1 mu /$}}#1}
\def\leftrightarrowfill{$\mathsurround=0pt \mathord\leftarrow \mkern-6mu
       \cleaders\hbox{$\mkern-2mu \mathord- \mkern-2mu$}\hfill
       \mkern-6mu \mathord\rightarrow$}
\def\dvec#1{\vbox{\ialign{##\crcr
       \leftrightarrowfill\crcr\noalign{\kern-1pt\nointerlineskip}
       $\hfil\displaystyle{#1}\hfil$\crcr}}}          
\def\hook#1{{\vrule height#1pt width0.4pt depth0pt}}
\def\leftrighthookfill#1{$\mathsurround=0pt \mathord\hook#1
       \hrulefill\mathord\hook#1$}
\def\underhook#1{\vtop{\ialign{##\crcr                 
       $\hfil\displaystyle{#1}\hfil$\crcr
       \noalign{\kern-1pt\nointerlineskip\vskip2pt}
       \leftrighthookfill5\crcr}}}
\def\smallunderhook#1{\vtop{\ialign{##\crcr      
       $\hfil\scriptstyle{#1}\hfil$\crcr
       \noalign{\kern-1pt\nointerlineskip\vskip2pt}
       \leftrighthookfill3\crcr}}}

\def\sfrac#1#2{{\vphantom1\smash{\lower.5ex\hbox{\small$#1$}}\over
       \vphantom1\smash{\raise.4ex\hbox{\small$#2$}}}} 
\def\bfrac#1#2{{\vphantom1\smash{\lower.5ex\hbox{$#1$}}\over
       \vphantom1\smash{\raise.3ex\hbox{$#2$}}}}      
\def\afrac#1#2{{\vphantom1\smash{\lower.5ex\hbox{$#1$}}\over#2}}  
\def\on#1#2{{\buildrel{\mkern2.5mu#1\mkern-2.5mu}\over{#2}}}
\def\ddt#1{\on{\hbox{\LARGE .\kern-2pt.}}#1}             
\def\tdt#1{\on{\hbox{\LARGE .\kern-2pt.\kern-2pt.}}#1}   

\def\boxes#1{
       \newcount\num
       \num=1
       \newdimen\downsy
       \downsy=-1.5ex
       \mskip-2.8mu
       \bo
       \loop
       \ifnum\num<#1
       \llap{\raise\num\downsy\hbox{$\bo$}}
       \advance\num by1
       \repeat}
\def\boxup#1#2{\newcount\numup
       \numup=#1
       \advance\numup by-1
       \newdimen\upsy
       \upsy=.75ex
       \mskip2.8mu
       \raise\numup\upsy\hbox{$#2$}}

\newskip\humongous \humongous=0pt plus 1000pt minus 1000pt
\def\caja{\mathsurround=0pt}
\def\eqalign#1{\,\vcenter{\openup2\jot \caja
       \ialign{\strut \hfil$\displaystyle{##}$&$
       \displaystyle{{}##}$\hfil\crcr#1\crcr}}\,}
\newif\ifdtup

\def\1ov4{{1\over 4}}

\def\pa{\partial}

\def\ddt{\dot{\t}}

\def\pa{\partial}

\def\pa{\partial}
\renewcommand{\a}{\alpha}
\renewcommand{\b}{\beta}

\renewcommand{\d}{\delta}
\newcommand{\rmd}{{\rm d}}

\newcommand{\beq}{\begin{equation}}
\newcommand{\eeq}{\end{equation}}

\def\ba{\begin{eqnarray}}
\def\ea{\end{eqnarray}}

\begin{document}
 



\null\vskip-24pt
\hfill KL-TH 00/03
\vskip-10pt
\hfill {\tt hep-th/0003017}
\vskip0.3truecm
\begin{center}
\vskip 3truecm
{\Large\bf
A note on $R$-currents and trace anomalies in the (2,0) tensor
multiplet in $d=6$ and AdS/CFT correspondence}\\ 
\vskip 1.5truecm
{\large\bf
Ruben Manvelyan 
 \footnote{On leave from Yerevan Physics Istitute, email:{\tt
manvel@physik.uni-kl.de}   
} and Anastasios C. Petkou
   \footnote{email:{\tt
       petkou@physik.uni-kl.de}}  
}\\
\vskip 1truecm
{\it Department of Physics, Theoretical Physics\\
University of Kaiserslautern, Postfach 3049 \\
67653 Kaiserslautern, Germany}\\

\end{center}
\vskip 3truecm
\centerline{\bf Abstract}

We discuss the  two- and three-point functions of the $SO(5)$
$R$-current of the (2,0) 
tensor multiplet in $d=6$, using AdS/CFT correspondence as well
as a free  
field realization. The results obtained via AdS/CFT correspondence
coincide with the ones from a free field calculation up to an overall $4N^3$
factor. This is the same factor found recently in studies of
the two- and
three-point functions of the energy momentum tensor in the (2,0)
theory. We connect 
our results to the trace anomaly in $d=6$ in the presence of external
vector fields and briefly discuss their implications. 
\newpage


Correlation functions of conserved currents carry
important information for a quantum field theory. In the case of
AdS/CFT correspondence \cite{mald}, their study has  unveiled
important properties of the 
boundary CFT. Particular examples are studies of the two- and
three-point functions of the energy momentum tensor
\cite{LT,gleb,BFT1,BFT2} and also of the two- and three-point functions of
conserved vector currents \cite{freed,chalm}. These
have provided important tests, both for the AdS/CFT correspondence and
also for the non-renormalization properties of ${\cal N}=4$ SYM in
$d=4$.
Recently, there has been a growing interest for the study of AdS/CFT
correspondence in diverse dimensions, such as AdS$_{4/7}$/CFT$_{3/6}$,
which are connected with M2/M5 branes respectively.
 The two- and three-point
functions of the energy momentum tensor have been recently discussed
in both the above cases \cite{BFT1,BFT2}. Such studies are of
particular interest as the 
boundary CFTs are not well understood. Our aim here is to extent
the results of \cite{BFT1,BFT2} for 
the (2,0) multiplet in $d=6$ by discussing the two- and three-point
functions of the $SO(5)$ $R$-current of the theory, both via
AdS$_7$/CFT$_6$ correspondence and also using a free field
realization. 

It is well
known \cite{ospet} that the general structure of two- and three-point
functions in a CFT is fixed by conformal invariance up to a number of
constants.  Let $J_{\mu}(x)=J^{A}_{\mu}(x) T^{A}$, $\mu=1,..,d$ be a
general conserved 
current. In the case when $J_{\mu}(x)$ corresponds to the $R$-current
of the boundary CFT$_d$ obtained via AdS$_{d+1}$/CFT$_d$
correspondence, $T^A$ are the Lie algebra generators of
the isometry group of the sphere resulting from   
the compactification of the corresponding M/IIB theory on $AdS_{d+1}\times
S^{(10/9-d)}$. Then, from conformal invariance we can write in the
(flat) boundary theory \cite{ospet}
\ba
\<J^{A}_{\m}(x)J^{B}_{\n}(y)\> &=& \d^{AB}\frac{{\cal
C}^{(d)}_{V}}{[(x-y)^2]^{(d-1)}}I_{\m\n}(x-y), \,\,\,\,I_{\m\n}(x)=\d_{\m\n}
-2\frac{x_{\m}x_{\n}}{x^{2}}\,, \label{corr1}\\  
\<J^{A}_{\m}(x)J^{B}_{\n}(y)J^{C}_{\l}(z)\> &=&
\frac{f^{ABC}}{(x-y)^{d-2}(x-z)^{d-2}(y-z)^{d}} \nonumber\\  
&&\times \, I_{\n\s}(x-y)\,I_{\l\r}(x-z)\,
I_{\s\b}(X)\,t_{\m\b\r}(X)\,,\label{corr2}\\  
X_{\m}&= &\frac{(x-y)_{\m}}{(x-y)^2} -
\frac{(x-z)_{\m}}{(x-z)^2} \,,\nonumber\\ 
t_{\m\n\l}(X) &=& {\cal A}^{(d)}\frac{X_{\m}X_{\n}X_{\l}}{X^{2}} + {\cal
B}^{(d)}\left( X_{\m}\d_{\n\l} + X_{\n}\d_{\m\l} -
X_{\l}\d_{\m\n}\right)\,,\nonumber 
\ea
where $f^{ABC}$ are the Lie algebra structure constants in the
representation carried by the currents.  The constants ${\cal C}^{(d)}_{V}$, ${\cal A}^{(d)}$ and ${\cal
B}^{(d)}$ are important 
parameters of the boundary CFT$_d$. We are interested both in the
strong coupling as well as 
in the free field theory  values of these parameters. The strong
coupling values
follow from AdS$_{d+1}$/CFT$_d$ correspondence \footnote{We consider
the Euclidean 
version  of 
AdS$_{d+1}$  space where  $\rmd \hat{x}^{\m}\rmd
\hat{x}_{\m}=\frac{1}{x_{0}^{2}} 
(\rmd x_{0}\rmd 
x_{0}+\rmd x^{i}\rmd x^{i})$, with $i=1,..,d.$ and
$\hat{x}_{\m}=(x_{0},x_{i})$. 
The boundary of this space is isomorphic
to {\bf{S}}$^{\rm d}$ since it consists of
{\bf{R}}$^{\rm d}$ at $x_{0}=0$ and a single point at $x_{0}=\infty$.}
and have been obtained in
\cite{freed,chalm}.  Namely, from the $d+1$ dimensional Lagrangian ${\cal
L}=-\frac{1}{4g^{2}_{SG_{d+1}}}\int \rmd^{d+1}\hat{x}
\,\sqrt{g}\,F^{A}_{ij}(\hat{x} )F^{A,ij}(\hat{x})$, $
i,j=0,..,d$ (without the anomalous Chern-Simons term), one
obtains
\ba
\label{c}
{\cal C}^{(d)}_{V}&=&\frac{\G(d)(d-2)}{2
\pi^{\frac12 d}\G(\frac12 d)}\frac{1}{g^{2}_{SG_{d+1}}}\,,\\ 
{\cal B}^{(d)}&=&\frac{(2d-3)\G(\frac12 d)}
{\pi^{\frac12 d}4(d-1)}\frac{{\cal C}_V^{(d)}}{g^{2}_{SG_{d+1}}}\,,\label{b}\\ 
\label{a}  
{\cal A}^{(d)}&=&\frac{d\,\G(\frac12 d)}{\pi^{\frac12 d}4(d-1)} \frac{{\cal
C}_V^{(d)}}{g^{2}_{SG_{d+1}}}\,. 
\ea    
These relations are in agreement with the general conformal Ward
identity \cite{ospet} 
\ba
{\cal C}_{V}^{(d)}&=&\left(\frac{1}{d}{\cal A}^{(d)} + {\cal
B}^{(d)}\right)S_{d}, 
\,\,\,\, S_{d}=\frac{2\pi^{\frac12 d}} 
{\G\left(\frac12 d\right)}\,.\label{ward}
\ea
The crucial strong coupling information provided by (\ref{c}),
(\ref{b}) and (\ref{a}) is, apart from the value of  
${\cal C}^{(d)}_{V}$, the {\it relative} scale
of ${\cal A}^{(d)}_{V}$ and ${\cal B}^{(d)}_{V}$ which is given by 
\ba {\cal A}^{(d)}=\frac{d}{2d-3}{\cal B}^{(d)}\,.
\label{ab} 
\ea 
A similar discussion has been presented in  
\cite{BFT1} for the parameters involved in the two-
and three-point functions of the energy momentum tensor. Setting
$d=4,6$ in (\ref{c})-(\ref{a}) we obtain 
\ba
{\cal C}^{(4)}_{V}&=& \frac{6}{\pi^{2}g^{2}_{SG_{5}}}
,\,\,\, {\cal A}^{(4)}=\frac{2}{\pi^{4}g^{2}_{SG_{5}}},\,\,\, 
{\cal B}^{(4)}=\frac{5}{2\pi^{4}g^{2}_{SG_{5}}}\,,\label{ads5num}\\ 
{\cal C}^{(6)}_{V}&=&\frac{120}{\pi^{3}g^{2}_{SG_{7}}},\,\,\,
{\cal A}^{(6)}= 
\frac{72}{\pi^{6}g^{2}_{SG_{7}}},\,\,\, {\cal B}^{(6)}=
\frac{108}{\pi^{6}g^{2}_{SG_{7}}}\,.\label{ads7num} 
\ea 
The values given in (\ref{ads5num}) correspond to the large-$N$,
large-$g_{YM}^2N$ limit (with $g_{YM}$ the gauge coupling), of the
${\cal 
N}=4$, $d=4$  
$SU(N)$ SYM, while the ones given in (\ref{ads7num}), presumably, to the
strong coupling limit of the (2,0), $d=6$ selfdual tensor multiplet. 

We can calculate the values of
the parameters above using a free field 
realization of the corresponding theories.  To obtain the
free field   
(one loop), contributions to these parameters  from general scalar and
fermion currents with a global symmetry group we follow \cite{ospet}
and define the composite 
vector current
\ba
J^{A}_{\m}&=&\fii T^{A}_{\fii}\partial_{\m}\fii + \bar{\j}
T^{A}_{\j}\g_{\m}\j\,,\label{freecurr}\\ 
\left[T^{A}_{\fii},T^{B}_{\fii}\right]&=&f^{ABC}T^{C}_{\fii},\,\,\,\,\,\,\,
\left[T^{A}_{\j},T^{B}_{\j}\right]=f^{ABC}T^{C}_{\j}\,,\label{freecurr1}\\
{\rm Tr}\left(T^{A}_{\fii}T^{B}_{\fii}\right)&=&-C_{\fii}\d^{AB},\,\,\,\,\,\,
{\rm Tr}\left(T^{A}_{\j}T^{B}_{\j}\right)=-C_{\j}\d^{AB}\,.\label{freecurr2}
\ea
Then, from \cite{ospet} we obtain in general
dimensions
\ba
{\cal C}_{V,free}^{(d)}&=&\left(\frac{C_{\fii}}{d-2}
+C_{\j}({\rm Tr}{\bf I}_{\j})\right)\frac{1}{S_{d}^2}\label{cfree}\,,\\
{\cal B}^{(d)}_{free}&=&\left(\frac{C_{\fii}}{2}\frac{1}{d-2}
+C_{\j}({\rm Tr}{\bf I}_{\j})\right)\frac{1}{S_{d}^3}\,,\label{bfree}\\
{\cal
A}^{(d)}_{free}&=&
\frac{C_{\fii}}{2}\frac{d}{d-2}\frac{1}{S_{d}^3}\,.\label{afree}  
\ea    
Here $({\rm Tr}{\bf I}_{\j})$ is the dimension of the spacetime spinors forming
the current  
(\ref{freecurr}). 
Note that the constants
(\ref{cfree})-(\ref{afree}) 
do not depend directly on the number of modes but rather on the
representations of 
the global symmetry group under which the scalars and fermions in the
theory transform. It is easy to see that the above free
field values satisfy  the conformal Ward identity
(\ref{ward}). Let us then try to reproduce the strong coupling AdS
results (\ref{ads5num}) and
(\ref{ads7num}) using the above free
field values. As we have mentioned, the important condition to be
satisfied is (\ref{ab}), since the Ward identity (\ref{ward}) implies
that the overall normalization of ${\cal C}^{(d)}_V$, ${\cal A}^{(d)}$
and ${\cal 
B}^{(d)}$ is the same. Substituting then
(\ref{bfree}) and (\ref{afree})  into (\ref{ab}) we obtain 
\ba
C_{\fii} = C_{\j}({\rm Tr}{\bf I}_{\j})\,.
\label{abfree}
\ea        
This can be interpreted as a selection rule for the free field multiplets.

Next, using (\ref{freecurr})-(\ref{afree})  we can easily evaluate the
free field values ${\cal C}^{(d)}_V$, ${\cal A}^{(d)}$ and ${\cal
B}^{(d)}$ for the cases of interest to us, namely ${\cal N}=4$
SYM in $d=4$ with 
$R$-symmetry  group  $SU(4)$ and 
$(2,0)$ selfdual tensor multiplet in $d=6$ with  $R$-symmetry
group $SO(5)$.  In both cases we have
contributions from scalars  and chiral fermions.\footnote{Note
that vector fields do not contribute to ${\cal C}^{(d)}_V$, ${\cal A}^{(d)}$
and ${\cal 
B}^{(d)}$ in $d=4$ \cite{ospet}.}
In the case of ${\cal N}=4$ SYM in $d=4$ the scalar 
bosons are in the 
antisymmetric {\bf 6} representation of $SU(4)$  and the chiral
fermions in the fundamental representation of $SU(4)$. We then obtain
\ba
C^{{\bf 6}}_{\fii}&=&1,\qquad C^{{\bf 4}}_{\j}=\frac{1}{2},\qquad
({\rm Tr}{\bf I}_{\j})=2\,,\label{n40}\\ 
{\cal C}^{(4)}_{V,free}&=&\frac{3}{8\pi^4},\qquad {\cal
B}^{(4)}_{free}=\frac{5}{32\pi^6},\qquad  {\cal
A}^{(d)}_{free}=\frac{1}{8\pi^6}\,.   
\label{n41} 
\ea
Note that the selection rule (\ref{abfree}) is satisfied. 

For
comparison with the results from AdS$_{5}$/CFT$_4$ correspondence 
we need the value of $1/g_{SG_5}^2$. This can be obtained for both cases of
interest here, $d+1=5,7$, considering the equations of motion of gauged
supergavity in 5 and 7 dimensions correspondingly in the presence of
non trivial scalar fields in the coset space $SL(n,{\bf
R})/SO(n)$. The $SO(n)$ group corresponds to the $R$-symmetry group of
the boundary CFT$_d$. Following
\cite{gleb2}, the maximally supersymmetric solution of the
equations of motion (with the scalar fields set to zero), corresponds
to an AdS$_{d+1}$ metric $g_{\mu\nu}$ with negative cosmological constant as
\beq
\label{sugra}
R_{\m\n}=\frac{4}{d-1}\,P\,g_{\m\n}\,,\,\,\,\,\,
P=-g^2\frac{n(n-2)}{32} \,,\,\,\,\,(n,d+1)=(6,5),\,(5,7)\,,
\eeq
where $g^2$ corresponds to the supergravity gauge coupling. Comparing
(\ref{sugra}) with the vacuum solution $R_{\m\n}=-d\,g_{\m\n}$ for
IIB/M theory  
compactifications on AdS$_{5/7}\times$S$^{5/4}$ with AdS radius one, we obtain
\beq
\label{new1}
g^2 = 8\frac{d(d-1)}{n(n-2)}\,.
\eeq
Taking then into account the relative 1/2 normalization between the
$R$ and $F^2$ terms in gauged supergravity \cite{gleb2,peter} we get
\beq
\label{new2}
\frac{1}{g_{SG_{d+1}}^2} =\frac{n(n-2)}{4d(d-1)} \frac{1}{2\kappa_{d+1}^2}\,.
\eeq
The value of the gravitational constant $1/2\kappa_{d+1}^2$ has been
obtained 
for $d+1=5,7$ in \cite{Klebanov,BFT1} as
\beq
\label{kappa}
\frac{1}{2\kappa_5^2}=\frac{N^2}{8\pi^2}\,,
\,\,\,\,\,\,\,\,\frac{1}{2\kappa_7^2}  =\frac{N^3}{3\pi^3}\,,
\eeq
where $N$ is the number of coincident $D3$/$M5$ branes.
By virtue of (\ref{kappa}) we then obtain for $d+1=5$, $n=6$   
\ba 
\frac{1}{g_{SG_5}^2} = \frac{N^2}{16\pi^2}\,.
\label{IIB}
\ea
This value
coincides with one obtained in \cite{freed,bilal}. Finally we obtain
\ba
\frac{{\cal C}^{(4)}_{V}}{{\cal C}^{(4)}_{V,free}}&=&\frac{{\cal
B}^{(4)}}{{\cal B}^{(4)}_{free}}=\frac{{\cal A}^{(4)}}{{\cal
A}^{(4)}_{free}}=N^2 \,.
\label{n2}
\ea 
The overall factor $N^2$ can be recognized as the large-$N$ value of
the 
dimension of the adjoint representation of $SU(N)$ carried by all fields
of ${\cal N}=4$  $SU(N)$ SYM theory. The same $N^2$ overall factor is
obtained in studies of the 
parameters involved in the two- and three-point functions of the
energy momentum tensor in ${\cal N}=4$ SYM in $d=4$ \cite{LT,gleb}.
The result (\ref{n2}) has been also obtained in 
\cite{freed,chalm}. It essentially implies  that the quantities
${\cal C}_V^{(4)}$, ${\cal A}^{(4)}$ and ${\cal 
B}^{(4)}$ are not renormalized as one goes from the weak (free fields) to
the strong coupling regime of the theory \cite{freed1}. 

Next we perform the same calculation for the $(2,0)$ supermultiplet
in $d=6$. 
The scalar fields are realized here as antisymmetric traceless
spin-tensors  in a spinor representation  of the $R$-symmetry group
$SO(5)$ \cite{toine}. Namely,
\ba 
\phi^{ij}=-\phi^{ji},\qquad \Omega_{ij}\phi^{ij}=0, \qquad i,j=1,2,3,4\,,
\label{phi}
\ea 
where  $\Omega_{ij}$ is the  symplectic ($\Omega^{T}=-\Omega$) invariant
tensor of $USp(4)$. 
The spinors of the $(2,0)$ supermultiplet form a symplectic
Majorana-Weyl spinor as
\ba
\j_{i}=\j^{j}\Omega_{ji}\,.
\label{psi}
\ea 
With these fields we construct the $R$-current with the scalars in the
${\bf 5}$ and the spinors in the ${\bf 4}$ of $SO(5)$ as follows
\footnote{We introduce $USp(4)$  $4\times 4$  gamma matrices 
$\g^{a}_{ij}$ and $\g^{ab}_{ij}=\frac{1}{2}\left[\g^{a},\g^{b}\right]_{ij}$} 
\ba
J^{ab}_{\m}&=&\phi^{ik}T^{ab}_{ik,jl}\partial_{\m}\phi^{jl} 
+ \bar{\j}^{i} t^{ab}_{ij}\g_{\m}\j^{j}, \qquad
J^{ab}_{\m}=-J^{ba}_{\m}\qquad a,b=1,2,..5\,,\label{cur1}\\ 
t^{ab}_{ij}&=&\frac{1}{4}\g^{ab}_{ij},\qquad  \qquad  \qquad
T^{ab}_{ik,jl} = \frac{1}{8}\left(\g^{ab}_{ij}\Omega_{kl}-
  \g^{ab}_{kj}\Omega_{il}
  +\g^{ab}_{kl}\Omega_{ij}-\g^{ab}_{il}\Omega_{kj}\right)\,,\label{cur2}\\     
C_{\j}^{\bf 4} &=& \frac{1}{2},\qquad  \qquad  \qquad  \qquad
C_{\phi}^{\bf 5}=1\,. 
\label{cur3}
\ea
The ratio $C^{\bf 5}_{\phi}/C^{\bf 4}_{\j}=2$ is fixed by the
respective representations of scalars and 
fermions. 

Using then 
(\ref{cfree})-(\ref{afree}) and  (\ref{phi})-(\ref{cur3}) we can
evaluate the parameters for the $(2,0)$,  
$d=6$ supermultiplet. For this we need to take $({\rm Tr}{\bf
  I}_{\psi})=2$ in (\ref{cfree})-(\ref{afree}) which amounts to
considering  half of the fermionic degrees of freedom
\cite{BFT1}. Note that with this prescription the 
selection rule (\ref{abfree}) is satisfied for the values of $C^{\bf 4}_{\j}$
and $C^{\bf 5}_{\phi}$ in 
(\ref{cur3}). We then obtain,    
\ba
{\cal C}^{(6)}_{V,free}&=&\frac{5}{4\pi^6},\qquad {\cal B}^{(6)}_{free}
=\frac{9}{8\pi^9},\qquad  {\cal A}^{(6)}_{free}=\frac{3}{4\pi^9}\,.
\label{2.0num}
\ea
We can compare the above free field values  with the
ones obtained via AdS$_{7}$/CFT$_6$ correspondence 
in (\ref{ads7num}), if we substitute 
from (\ref{new2}) the corresponding value for $1/g^2_{SG_7}$ 
\ba 
\frac{1}{g_{SG_7}^2} = \frac{N^3}{24\pi^3}\,.
\label{g7}
\ea 
By virtue of
(\ref{g7}) we obtain from (\ref{ads7num}) and (\ref{2.0num}) 
\ba
\frac{{\cal C}^{(6)}_{V}}{{\cal C}^{(6)}_{V,free}}&=&\frac{{\cal
B}^{(6)}}{{\cal B}^{(6)}_{free}}=\frac{{\cal
A}^{(6)}}{{\cal A}^{(6)}_{free}}=4N^3 \,.
\label{n3}
\ea 
The overall factor $4N^3$ is the same with the one found 
in studies of the two- and three-point functions of the energy momentum
tensor \cite{BFT1}. 

An important property of the parameters which appear in two- and
three-point functions of conserved currents (in even dimensions), is
that they are 
intimately connected to the trace anomaly of the theory in the
presence of external sources. For the (2,0) multiplet in $d=6$, such
external sources for the $R$-current are 
introduced if one adds to the action the term
\ba
\int\rmd^{d}x\,\sqrt{g}\,g^{\m\n}A^{ab}_{\m}(x)J^{ab}_{\n}(x)\,.
\label{aj}
\ea
Conservation of the $R$-current  means that the external field is
defined up to a gauge transformation $A^{ab}_{\m}(x)\sim A^{ab}_{\m}(x)
+ \partial_{\mu}\a^{ab}(x)$. The $R$-current is in the 
supermultiplet of conserved currents, which also includes the energy
momentum tensor \cite{toine}. Therefore, 
we can consider a supersymmetric introduction of the external
source $A^{ab}_{\m}(x)$, if we view the latter as a component of 
a general $d=6$ background superfield which also includes an external
gravitational field. Similar considerations were presented
in \cite{freed1} in the case of four-dimensional theories. Then, a
simple extension of  
the results in \cite{ospet} allows us to connect all three constants
${\cal C}^{(d)}_V$, ${\cal A}^{(d)}$ and ${\cal  
B}^{(d)}$ to the trace anomaly of the (2,0) theory in $d=6$ in the presence
of external vector fields. The general idea underlying such a
connection is that the external trace anomaly is tied to the
short distance singularities of the renormalized $n$-point functions. In
the case of interest here,  we can follow \cite{petsken} and write
the general 
renormalization group equation 
\ba
&&\sum_{k=1}^{\infty}\frac{1}{k!} \int\rmd^{6}x_{1}\sqrt{g}g^{\m_{1}\n_{1}} 
\,..\,\rmd^{6}x_{k}\sqrt{g}g^{\m_{k}\n_{k}}
A^{a_{1}b_{1}}_{\m_{1}}(x_{1})..
A^{a_{k}b_{k}}_{\m_{k}}(x_{k})\,\m\frac{\partial}{\partial\mu}
\langle J^{a_{1}b_{1}}_{\n_{1}}(x_{1})..J^{a_{k}b_{k}}_{\n_{k}}(x_{k})
\rangle_{R}\nonumber\\ &&\hspace{1cm}=\int\rmd^{6}x\sqrt{g}g^{\m\n}(x)\langle
T_{\m\n}(x)\rangle\,.\label{rg}
\ea 
The
subscript $R$ in the first line of (\ref{rg}) denotes the renormalized
$n$-point functions which 
depend on the arbitrary mass parameter $\m$. Taking suitable
functional derivatives of (\ref{rg}) with respect to $A_{\m}^{ab}(x)$
we can, in principle, connect the parameters which appear in the two-
and three-point functions of $J_{\mu}^{ab}(x)$ with the possible terms
in the trace anomaly. 

The general structure of the conformal
anomaly in $d=6$ in the presence of 
external vector fields can be obtained following considerations similar
to the ones used in the 
classification of 
conformal anomalies in the presence external gravitational fields
\cite{anom}. In the flat space limit and up
to total derivatives  
there exist only {\it two} independent gauge and scale
invariant contributions to the external trace anomaly. Here we choose
to parametrize the anomaly as 
\ba
\langle T_{\m}^{\,\,\m}(x)\rangle =
\alpha_{V}\,F^{ab,\,\n}_{\m}F^{cd,\,\l}_{\n}F^{pq,\,\m}_{\l}f^{ab,cd,pq}
+  
\beta_{V}\,\partial^{\m}F^{ab}_{\m\n}\,\partial^{\l}F^{ab,\,\n}_{\l}\,.
\label{t}
\ea
It is easy to see that only the second term on the r.h.s. of (\ref{t})
contributes to the
two-point function in (\ref{rg}). To proceed we need the
renormalized, $\mu$-dependent expression for the two-point function
(\ref{corr1}) in $d=6$. This is equivalent to obtaining the
renormalized expression for the non-integrable singularity $1/x^8$ as
$x\rightarrow 0$. The latter is accomplished by the substitution
$1/x^8 \rightarrow \car \left(1/x^8\right)$ -  which corresponds to
real-space renormalization \cite{freed2,ospet} - and the use of the
general formula  
\cite{petsken}
\ba
\m {\pa \over \pa \m} \left[\car \left({1 \over (x^2)^{{\frac{1}{2}d}
+k}}\right)\right] =  
\frac{\Gamma(\frac{1}{2}d)}{4^{k}\Gamma(k+1)
\Gamma(\frac{1}{2}d+k)} S_{d} (\pa^2)^k \d^{(d)}(x)\,.
\label{ren}
\ea 
Then, from   (\ref{corr1}), (\ref{rg}) and (\ref{t}) we obtain by
virtue of (\ref{ren}) 
\ba
\beta_{V} = \frac{{\cal C}^{(6)}_{V}\pi^3}{960} =  \frac{N^3}{192\pi^3}\,,
\label{betafinal}
\ea 
where to get the last equality we have used
(\ref{ads7num}) and (\ref{g7}). The result (\ref{betafinal}) combined
with (\ref{n3}) implies that the trace anomaly coefficient $\b_V$ is
non-renormalized as we go from the weak (free fields) to the strong
coupling regime of the (2,0) multiplet in $d=6$ (up to an overall
factor $4N^3$). 

Following the same reasoning as above, one expects on general grounds that 
the properly renormalized three-point function (\ref{corr2})
involves certain additional ultralocal structures (e.g. products of
delta functions), which are proportional to the two parameters ${\cal
A}^{(6)}$ and 
${\cal B}^{(6)}$. This would lead, through
(\ref{rg}), to a certain {\it linear} relationship between ${\cal A}^{(6)}$, 
${\cal B}^{(6)}$ and the coefficient of the conformal anomaly
$\a_V$.\footnote{Note that only the first term on the r.h.s. of
(\ref{t}) contributes to the three-point function in
(\ref{rg}). Similar arguments have been used to connect the 
external trace 
anomalies to the parameters in the two- and three-point functions of
the energy momentum tensor \cite{ospet,BFT2}.} A proper
analysis of the short-distance singularities in 
three-point functions will not be attempted here as it is
significantly more complicated and has only be achieved in simple
cases \cite{ospet}. Nevertheless, the above arguments imply that the
expected relationship between ${\cal A}^{(6)}$, 
${\cal B}^{(6)}$ and $a_V$, which would follow from such an
analysis, should hold both in the strong and the weak coupling
regimes of the relevant CFT. If then, as we have shown in (\ref{n3}),
the {\it relative} scale between ${\cal A}^{(6)}$ and ${\cal B}^{(6)}$
is the same both in the strong and the weak coupling regimes of the (2,0)
multiplet in $d=6$, we conclude that the coefficient $\a_V$
remains also the same in both the strong and the weak coupling regimes
of the theory (up
to the overall factor $4N^3$).

Concluding, we have shown the agreement of the free field theory and
AdS$_7$/CFT$_6$ results for the parameters appearing in the two- and
three-point functions of the 
$R$-symmetry current of the (2,0) multiplet in $d=6$, up to an
overall $4N^3$ factor. The same factor was found in studies of
the two- and 
three-point functions of the energy momentum tensor. Moreover, we
argued that such a result implies that the coefficients of the trace 
anomaly  in the
presence of external sources for the $R$-currents are non-renormalized
up to the same overall $4N^3$ factor. The trace anomaly is in the same
supermultiplet as the $R$-current anomaly. In the case of ${\cal N}=4$
SYM in $d=4$, the Adler-Bardeen theorem protects the latter from being
renormalized leading to a non-renormalization theorem for the external
trace anomaly \cite{freed1}. In the case of the (2,0) multiplet in
$d=6$, it was found in \cite{BFT2} that some part of the external
trace anomaly - the part proportional to the
six-dimensional Euler density - {\it does} renormalize as one goes from the
weak to the strong coupling regime. In view of such a result, a better
understanding of the renormalization properties of the (2,0) multiplet
in $d=6$ is required. This may be achieved by studies of the
$R$-current anomaly (see \cite{intri} for a recent work), using also the
approach developed here. Studies of the $R$-currents
in  
CFTs obtained via AdS$_4$/CFT$_3$ correspondence are also of great
interest, however the absence of a trace anomaly there might lead to
less transparent results.   

\section*{Acknowledgements}

We would like to thank G. Arutyunov for valuable
discussions. We would also like to thank K. Intriligator for his very
illuminating remarks after the first version of the manuscript.
R. M. and A. C. P. are supported by the
Alexander von Humboldt Foundation.

\end{document}